\documentclass[twocolumn,aps,superscriptaddress]{revtex4}
\usepackage{amssymb}
\usepackage{amsmath}
\usepackage{graphicx}
\usepackage[normalem]{ulem}
\usepackage[dvips]{color}
\usepackage{lineno}

\renewcommand\sout{\bgroup \color{red} \ULdepth=-.5ex \ULset}

\begin{document}

%\preprint{}
\title{Effects of high-momentum tail of nucleon momentum distribution on initiation of cluster production in heavy-ion collisions at intermediate energies}
\author{Fang Zhang \footnote{zhangfang@lzu.edu.cn}}
\affiliation{School of Nuclear Science and Technology, Lanzhou University}
\author{Gao-Chan Yong \footnote{yonggaochan@impcas.ac.cn}}
\affiliation{Institute of Modern Physics, Chinese Academy of
Sciences, Lanzhou 730000, China}
\affiliation{School of Nuclear Science and Technology, University of Chinese Academy of Sciences, Beijing 100049, China}

\date{\today}

\begin{abstract}

Based on the transport model isospin-dependent Boltzmann-Uehling-Uhlenbeck coupled with a phase-space coalescence after-burner, we studied the effects of the high-momentum tail (HMT) of nucleon momentum distribution in initialization in $^{197}Au$+$^{197}Au$ reactions at a beam energy of 400 MeV/nucleon with different impact parameters. We found remarkable impact parameter dependent HMT effects on the fragment multiplicity distribution. The average neutron to proton ratio of produced isotopes is also affected by the HMT. The rapidity distributions of triton and $^3$He and their elliptic flows are all evidently affected by the HMT. All the effects of the HMT on the cluster production in heavy-ion collisions are centrality dependent.

\end{abstract}

%\pacs{25.70.-z, 25.70.Pq, 21.65.Ef}
\maketitle

\section{Introduction}

The high-momentum transfer experiments have shown that nucleons in
nuclear ground states can form short range correlated (SRC) pairs with center-of -mass  momentum lower than nuclear Fermi momentum ($<k_{F}$) and larger relative momentum ($>k_{F}$)\cite{prl04,prl05,prl06,prl07} due to the short-range nucleon-nucleon tensor interactions \cite{M05,Schiavilla07,Rios14,XC13}. The momentum distribution from low momenta ($k<k_{F}$) to high momenta ($k>k_{F}$) presents a high momentum tail (HMT) for nucleons. Furthermore, two-nucleon knock-out reactions have identified that nucleons in the HMT are isospin dependent \cite{sci08,sci14}, i.e., the $p-n$ SRC pairs are about 18 times that of $n-n$ SRC pairs or $p-p$ SRC pairs, referred to as `` $np$ dominance ''. The recent high-energy electron knockout experiments conducted by the CLAS Collaboration using medium-to-heavy nuclei show that the fraction of the high-momentum protons increases significantly with the neutron excess in nucleus whereas the fraction of high-momentum neutrons decreases slightly \cite{clasnature18}. These results would have significant implications for understanding the dynamics of neutron-rich reaction and the physics of neutron stars \cite{li18}.

It is well known that in intermediate-energy heavy-ion collisions (HICs), both nucleonic mean-field and nucleon-nucleon collisions are at work. The frequently mentioned phenomenon in intermediate-energy HICs is multi-fragmentation and the production of fragments \cite{liu01,Zhang99}. As the HMT of the single-nucleon momentum distribution in the nucleus is confirmed by experiments, the HMT has been considered in a certain transport model \cite{yong20151,yong20152}, and some isospin-sensitive observables, such as the nucleonic collective flows, the free neutron to proton ratio $n/p$ as well as the $\pi^{-}/\pi^{+}$ ratio. They all have been shown to be more or less sensitive to the HMT \cite {cpl16,PJA16}. Since the productions of fragments and the light clusters, e.g., triton and $^3$He are frequently studied in the literature \cite{yongflow09,lee}, in the present paper, we investigate the HMT effects on the production of fragments and the light clusters triton ($t$) and $^3$He in the $^{197}\rm {Au}+^{197}\rm{Au}$ collisions at the incident beam energy 400 MeV/nucleon with various collision impact parameters.

\section{The transport model and the nucleon momentum initialization}

In the present paper, we adopt the semiclassical isospin-dependent Boltzmann-Uehling-Uhlenbeck (IBUU) transport model and the phase-space coalescence afterburner \cite{yongflow09}. The initial density distributions of nucleons in the projectile and target are obtained from the Skyrme-Hartree-Fork calculations with the Skyrme $M^{*}$ parameters \cite{jfprc86}.
We use a nucleon momentum distribution with a high-momentum tail reaching $\lambda k_{F_{n,p}}=2.2 k_{F_{n,p}}$, where $\lambda=k_{max}/k_{F}$ is the high momentum cutoff parameter and $k_{F_{n,p}}$ is the neutron or proton Fermi momentum \cite{henprc15,yong20152}. According to the $n$-$p$ dominance model\cite{sci08,sci14}, $20\%$ of nucleons with equal numbers of neutrons and protons are inside the HMT. The momentum distribution for nucleons with high-momentum tail in initialization reads
\begin{eqnarray}
n^{HMT}(k) \propto 1/k^{4}
\end{eqnarray}
and
\begin{eqnarray}
n(k)=\left\{%
  \begin{array}{ll}
    C_{1}, & \hbox{$k \leq k_{F}$;} \\
    C_{2}/k^{4}, & \hbox{$k_{F} < k < \lambda k_{F}$}, \\
\end{array}%
\right.
\label{nk}
\end{eqnarray}
keeping 20\% fraction of total nucleons in the HMT, i.e.,
\begin{equation}
\int_{k_{F}}^{\lambda k_{F}}n^{HMT}(k)k^{2}dk \bigg/ \int_{0}^{\lambda k_{F}}n(k)k^{2}dk = 20\%
\end{equation}
with the normalization condition
\begin{equation}
\int_{0}^{\lambda k_{F}}n(k)k^{2}dk = 1.
\end{equation}
In the above equations, the parameters $C_{1}$ and $C_{2}$ in Eq.~(\ref{nk}) are determined automatically from the above equations.
With the above nucleon momentum distribution form, the average kinetic energy of nucleons increases roughly several MeV comparing to that with the ideal Fermi-gas approximation (in this case, nucleon momentum distribution $n(k)$ ranges in 0-$p_{F}$) without considering the HMT at 400MeV/nucleon beam energy \cite{yongplb17}. For neutron-rich nuclei, the proton has a larger probability than the neutron with momentum greater than the nuclear Fermi momentum \cite{yongprc17}. For example, there are about $25\% (17\%)$ protons (neutrons) with momenta larger than the proton (neutron) Fermi momentum in $^{197}Au$ nucleus \cite{yongplb17}. This phenomenon may lead to a result that protons have higher average momentum than neutrons in neutron-rich nuclei \cite{sci14}. The evolution of the nucleon momentum distribution in nucleus $^{197}Au$ with and without the HMT has also been studied in the previous work \cite{yong20152}. The results showed that the shape of the initial momentum distribution changes insignificantly for nuclear collisions at a beam energy of 400 MeV per nucleon.

The single nucleon potential is one of the most important inputs for the BUU-like transport model in intermediate energies. In this paper, the momentum- and isospin- dependent single nucleon potential derived from the Hartree-Fock approach using a modified Gogny effective interaction \cite{das03,yong18} is employed. Especially, the single particle potential of a nucleon with isospin $\tau$ and momentum $\vec{p}$ at total density $\rho$ reads
\begin{eqnarray}
U(\rho,\delta,\vec{p},\tau)&=&A_u(x)\frac{\rho_{\tau'}}{\rho_0}+A_l(x)\frac{\rho_{\tau}}{\rho_0}\nonumber\\
& &+B(\frac{\rho}{\rho_0})^{\sigma}(1-x\delta^2)-8x\tau\frac{B}{\sigma+1}\frac{\rho^{\sigma-1}}{\rho_0^\sigma}\delta\rho_{\tau'}\nonumber\\
& &+\frac{2C_{\tau,\tau}}{\rho_0}\int
d^3\,\vec{p^{'}}\frac{f_\tau(\vec{r},\vec{p^{'}})}{1+(\vec{p}-\vec{p^{'}})^2/\Lambda^2}\nonumber\\
& &+\frac{2C_{\tau,\tau'}}{\rho_0}\int
d^3\,\vec{p^{'}}\frac{f_{\tau'}(\vec{r},\vec{p^{'}})}{1+(\vec{p}-\vec{p^{'}})^2/\Lambda^2},
\label{buupotential}
\end{eqnarray}
The single particle potential $U(\rho,\delta,\vec{p},\tau)$ with isospin $\tau$ = 1/2 (-1/2) is for neutron (proton). Here $\rho_0$ denotes saturation density, $\rho_{\tau}$ and $\rho_{\tau'}$ denote neutron or proton density with $\tau \neq \tau'$. $\delta=(\rho_{n}-\rho_{p})/(\rho_{n}+\rho_{p})$ is the isospin asymmetry. $C_{\tau,\tau}$ and $C_{\tau,\tau'}$ are the interaction strength parameters, which describe the momentum-dependent interaction for a nucleon of isospin $\tau$ with like and unlike nucleons in the background fields. For nucleon-nucleon collisions, the reduced in-medium $N N$ cross section is used. More details about the above single nucleon potential and the baryon-baryon cross section can be found in Ref. \cite{yongp4}.

Because most BUU-type transport models are incapable of forming dynamically realistic nuclear fragments, we use the phase-space coalescence model as an afterburner, see, e.g., refs. \cite{chen98,Zhang99,yongflow09}. In this coalescence model, a physical fragment is considered as nucleons with relative momenta smaller than $P_{0}$ and relative distances short than $R_{0}$. In our simulations, we use $P_{0}=263$  MeV/c and $R_{0}=3$ fm \cite{yongflow09}, but the qualitative results studied here are not sensitive to the above two parameters.

\section{Results and discussions}

\begin{figure}[thb]
\begin{center}
\includegraphics[width=0.45\textwidth]{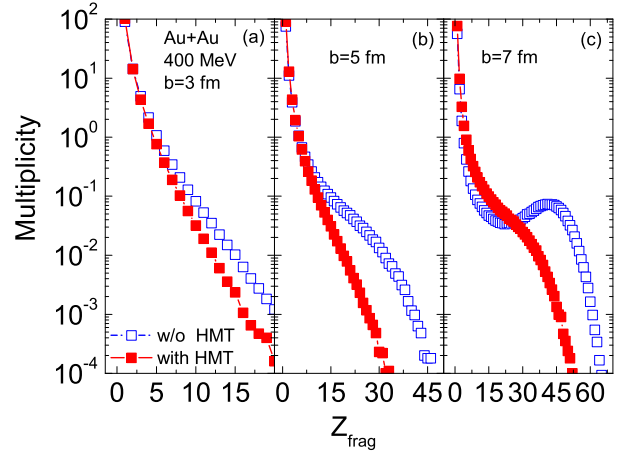}
\end{center}
\caption{Effects of the HMT on the multiplicities of fragments as a function of fragment charge in the Au+Au reactions at 400 MeV/nucleon with different impact parameters.} \label{multi}
\end{figure}
\begin{figure}[thb]
\begin{center}
\includegraphics[width=0.45\textwidth]{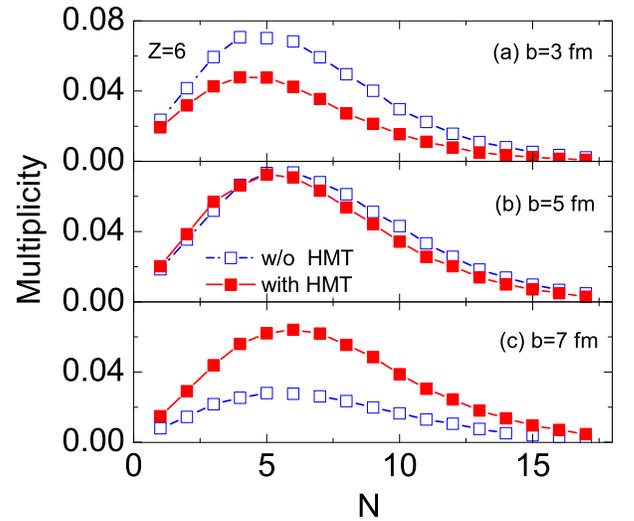}
\end{center}
\caption{Effects of the HMT on the yields of the isotope with charge number $Z$ = 6 as a function of neutron number in the Au+Au reactions at 400 MeV/nucleon with different impact parameters.} \label{z6}
\end{figure}
Our results presented here are obtained using totally 100,000 events in each case with 250 test particles per nucleon in each run of the simulation. To study the HMT effects on the fragment production in heavy-ion collisions, it is necessary to first show the effects of the HMT on the whole range of the fragments produced in HICs. Figure.\ \ref{multi} shows multiplicity distributions of fragments produced in HICs with and without the HMT. It is seen that the yields of fragments decrease with the increase in the fragment charge whether with or without the HMT. It is known that the violence of tge reaction decreases with the decrease in participant zone, so the effects of the HMT are more prominent in the collisions with larger impact parameters. The yields of very light fragments slightly increase with the HMT with different centralities, especially for larger centrality. For heavier fragments, the situation is inverse, i.e., the yields of heavier fragments evidently decrease with the HMT, especially with larger impact parameters. This is because the very light fragments are mostly pre-equilibrium emissions, the momentum correlations among nucleons increase the coalescence probability
among nucleons, thus more light fragments are produced. The heavier fragments are mainly deexcited remnants of the reaction, thus, heavier fragments are produced with smaller nucleon momenta (i.e., without the HMT). Therefore, for a specific fragment, such as the isotope with charge number $Z$ = 6 as shown in Fig.\ \ref{z6}, the HMT can increase or decrease its yields due to different centralities of the reaction.

\begin{figure}[thb]
\begin{center}
\includegraphics[width=0.45\textwidth]{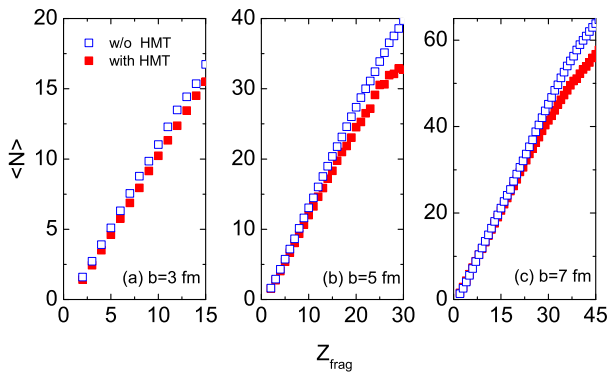}
\end{center}
\caption{Effects of the HMT on the average neutron number $<N>$ as a function of charge number $Z_{frag}$ in the Au+Au reactions at 400 MeV/nucleon with different impact parameters.} \label{aveN}
\end{figure}
\begin{figure}[thb]
\begin{center}
\includegraphics[width=0.45\textwidth]{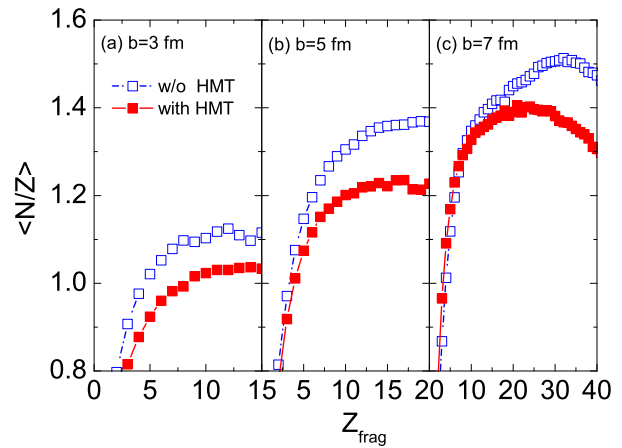}
\end{center}
\caption{Effects of the HMT on the average neutron to proton ratio $<N/Z>$ as a function of charge number $Z_{frag}$ in the Au+Au reactions at 400 MeV/nucleon with different impact parameters.} \label{RNZ}
\end{figure}
Since there is a predominance of isospin dependence in short-range correlated pairs, in addition to the multiplicity distribution,
the effects of the HMT should be also shown in average neutron number
$<N>$ and neutron to proton ratio $<N/Z>$ of the fragments. In Figs.\ \ref{aveN} and .\ \ref{RNZ}, we display the $<N>$ and $<N/Z>$ as a function of the fragment charge number ($Z_{frag}$) for the Au+Au reactions at the incident beam energy of 400 MeV/nucleon with and without the HMT in central
(left panel), semicentral (middle panel) and semiperipheral collisions (right panel).
From Fig.\ \ref{aveN}, one can see that with the HMT the average neutron number
$<N>$ decreases for heavier isotopes. This is understandable since the HMT causes de-excitation of remnants easily and usually more neutrons are emitted for neutron-rich system. To show more clearly the
isospin effects of produced isotopes, we analyze the
average neutron to proton ratio $<N/Z>$ of produced isotopes, which is defined as \cite{shetty03}
\begin{equation}
<N/Z>=\frac{\sum Y(X^{A}_{Z})(A-Z)/Z}{\sum Y(X^{A}_{Z})},
\end{equation}
where $Y(X^{A}_{Z})$ denotes the yield of the isotope $^{A}X$ of element $X$, the
summation is over all different isotopes.
The $Z_{frag}$ distribution of the $<N/Z>$ with and without the HMT is examined in Fig.\ \ref{RNZ}.
It is seen that the $Z_{frag}$ distribution of average
neutron to proton ratio $<N/Z>$ is sensitive to
the HMT especially for heavier fragments. The fact that the HMT decreases the
$<N/Z>$ of isotopes is consistent with that shown in Fig.\ \ref{aveN}.
Since heavier fragments are produced for the semiperipheral collisions,
the $<N/Z>$ ratio increases (approaching the reaction system's neutron to proton ratio $N/Z$ = 1.49) for a larger impact parameter.

\begin{figure}[thb]
\begin{center}
\includegraphics[width=0.4\textwidth]{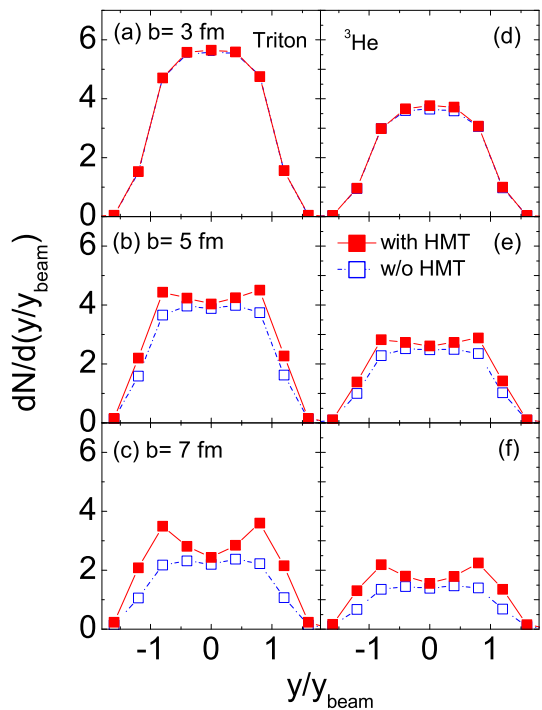}
\end{center}
\caption{Effects of the HMT on the rapidity distributions of triton and $^{3}$He in the Au+Au reactions at 400 MeV/nucleon with different impact parameters.} \label{dNdy}
\end{figure}
We next turn to study the effects of the HMT on light cluster production. Figure.\ \ref{dNdy} shows the rapidity distribution of triton and $^{3}$He in the $^{197}$Au+$^{197}$Au reaction at the incident energy of 400 MeV/nucleon with different impact parameters. It is seen that these clusters are mostly produced at midrapidities and their yields decrease with the increase in impact parameter \cite{chen032}. Because the HMT trends to deexcite remnants in HICs, more free nucleons and light clusters are, thus, produced and the free nucleons could also coalesce into light clusters. Therefore the HMT increases the yields of triton and $^{3}$He especially for larger impact parameters. At larger centralities, the effects of the HMT are more prominent around projectile and target rapidities.

\begin{figure}[thb]
\begin{center}
\includegraphics[width=0.45\textwidth]{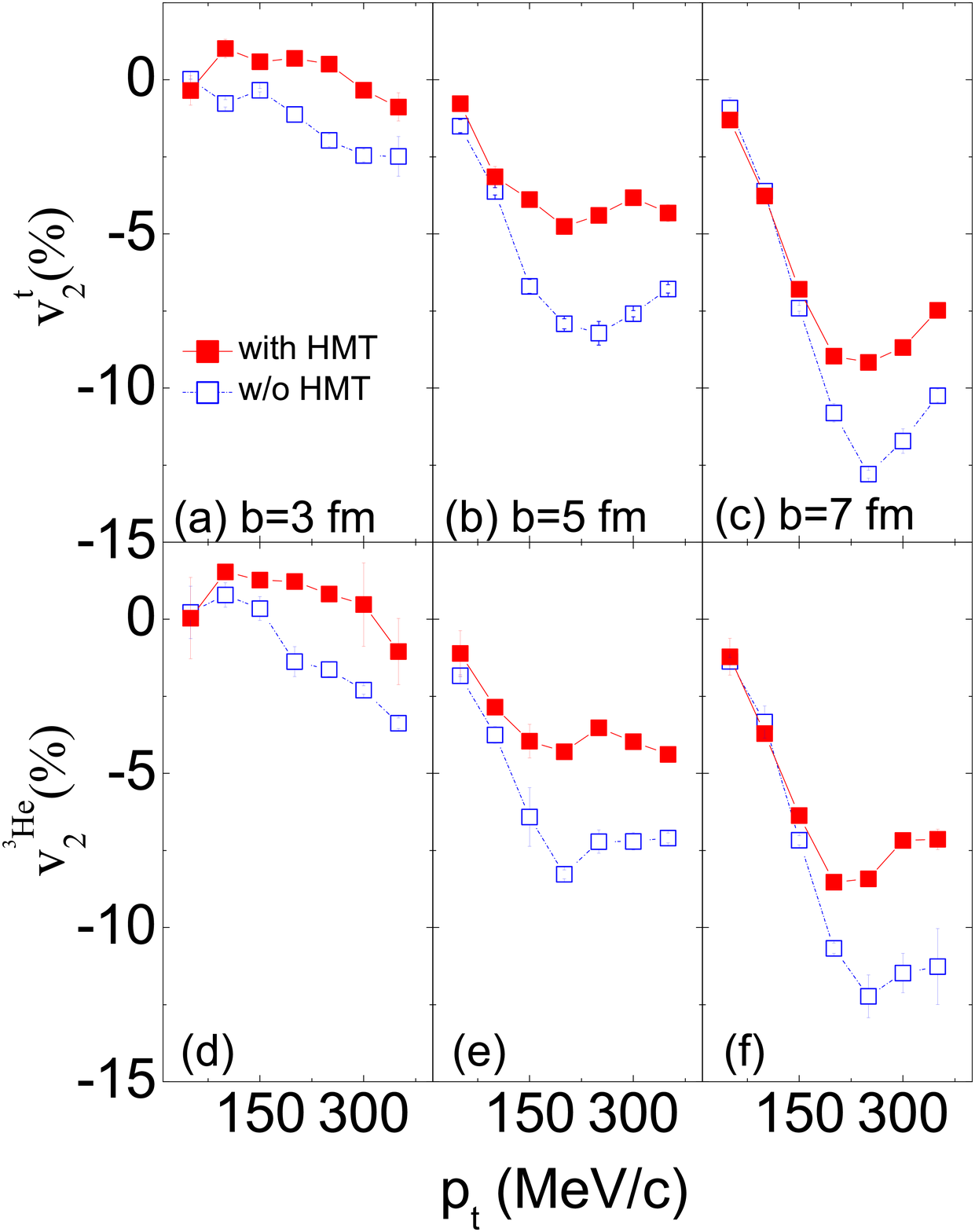}
\end{center}
\caption{Effects of the HMT on the elliptic flows of triton and $^{3}$He as a function of transverse momentum in the Au+Au reactions at 400 MeV/nucleon with different impact parameters.} \label{v2flow}
\end{figure}
Elliptic flow is related to the second Fourier coefficient of the azimuthal distribution, which reflects the anisotropy in the particle transverse momentum distribution at midrapidity can be expressed as \cite{lib99,gao18,ma01,zhfs01,ditorof}
\begin{equation}\label{}
  v_{2}=<cos(2\phi)>=\langle \frac{p_{x}^{2}-p_{y}^{2}}{p_{x}^{2}+p_{y}^{2}}\rangle.
\end{equation}
In the above, $p_{x}$ and $p_{y}$ are the in-plane and out-of-plane transverse momenta, respectively. The sign and magnitude of $v_{2}$ reflect the result of competition between the early out-of-plane nucleon emission and the later in-plane nucleon emission. $v_{2}>0$ indicates in-plane flow dominance, and $v_{2}<0$ describes out-of-plane flow dominance.
Obviously, $v_{2}=0$ reflects the isotropic distribution in the transverse plane.

From Fig.\ \ref{v2flow}, it is seen that the transverse momentum distributions of triton and $^{3}$He elliptic flows become flat in central collisions. This is because most of
nucleons participate in the reaction for central collisions, the emission of the particles tends to be isotropic. The size of spectators increases with the impact parameter, as a result, in semicentral and semiperipheral collisions we find preferential out-of-plane particle emissions, i.e., the squeezed-out effects of nucleon emission perpendicular to the reaction plane. An obvious HMT effect is seen on both triton and $^{3}$He elliptic flows at high momenta especially in semicentral and semiperipheral collisions.
%The absolute values of the triton and $^{3}$He elliptic flows become smaller in the cases with the HMT, %which can be understood from the following reasons. One reason is the fact that high $p_{t}$ particles %can only be produced through the most violent collisions in the early stage of the reaction %\cite{liprc01}. A larger early pressure gradients is created in the participant region when the particles %have higher average nucleon momenta \cite{jy92,jy98}, and it leads to a weaker ''squeeze out'', this %feature is revealed more clearly by the value of $v_{2}$ at high transverse momenta. The other reason %derives from
Since nucleonic correlations in momentum space decreases anisotropy of particle emission in HICs \cite{PJA16}, it is not surprising to see the strength of the elliptic flow with the HMT is smaller than without the HMT.
%The results in Fig.\ \ref{v2flow} indicate that nucleon-nucleon collisions involving HMT nucleons play an %important role for triton and $^{3}$He light clusters with transverse momenta higher than 250 MeV/c.

%Since there are only about $20\%$ nucleons in short range correlations, the experimental measurements of %HMT effects can easily influenced by the experimental conditions. In order to reduce the influence the %uncertainties of measurements, larger isospin asymmetry colliding nuclei can be chosen to study the HMT %effects according to n-p dominance in SRC. Then how to verify wether the experimental results actually %have HMT effect?  For this question, may be experimental identification can be started from where the HMT %effect is particularly pronounced, e.g., for triton and $^{3}$He elliptic flows with larger centralities %and transverse momenta above 250 MeV/c, and made theoretical simulation calculations for verification.

To confirm the effects of the HMT on the cluster production in heavy-ion collisions and check the model reliability, it is meaningful to carry out relevant experimental measurements, such as the multiplicity of fragments and the average neutron to proton ratio as a function of the charge number as well as the elliptic flows of the triton and $^{3}$He in Au+Au reactions at 400 MeV/nucleon. By comparison of theoretical predictions and experimental measurements, one can get new insights into the effects of nucleon-nucleon short-range correlations on the cluster production in heavy-ion collisions.

\section{Summary}

Based on a hybrid approach of the semiclassical transport model
IBUU coupling a phase-space coalescence after-burner, we studied the HMT effects on some
fragment observables in $^{197}Au+^{197}Au$ reactions at the incident energy of $400$
MeV/nucleon with impact parameters $b$=3, 5, and 7 fm, respectively. We found that distributions of the fragment multiplicity, the average neutron number of isotopes, as well as the average neutron to proton ratio of isotopes are all affected by the HMT. For the light clusters triton and $^{3}$He, both their multiplicities and their elliptic flows show clear effects of the HMT. Also the effects of the HMT on the cluster production are affected by the collision centrality in HICs. Therefore, the high-momentum tail of the nucleon momentum distribution in initialization deserves further studies whereas studying light and heavier cluster productions in heavy-ion collisions at low and intermediate energies.

%\begin{acknowledgments}

This work was supported by the National Natural Science Foundation of China under Grant No. 12275322 and
the Strategic Priority Research Program of Chinese Academy of Sciences with Grant No. XDB34030000.

%\end{acknowledgments}


\begin{thebibliography}{99}

\bibitem{prl04}R. A. Niyazov \emph{et al.}, Phys. Rev. Lett. {\bf 92}, 052303 (2004).
\bibitem{prl05}F. Benmokhtar \emph{et al.}, Phys. Rev. Lett. {\bf 94}, 082305 (2005).
\bibitem{prl06}K. S. Egiyan \emph{et al.}, Phys. Rev. Lett. {\bf 96}, 082501 (2006).
\bibitem{prl07}R. Shneor \emph{et al.}, Phys. Rev. Lett. {\bf 99}, 072501 (2007).
\bibitem{M05}M. M. Sargsian, T. V. Abrahamyan, M. I. Strikman and L. L. Frankfurt, Phys. Rev. C {\bf 71}, 044615 (2005).
\bibitem{Schiavilla07}R. Schiavilla, R. B. Wiringa, S. C. Pieper and J. Carlson, Phys. Rev. Lett. {\bf 98}, 132501 (2007).
\bibitem{Rios14}A. Rios, A. Polls and W. H. Dickhoff, Phys. Rev. C {\bf 89}, 044303 (2014).
\bibitem{XC13}C. Xu and B. A. Li, arXiv:1104.2075.
\bibitem{sci08}R. Subedi \emph{et al.}, Science {\bf 320}, 1476 (2008).
\bibitem{sci14}O. Hen \emph{et al.}, Science {\bf 346}, 614 (2014).
\bibitem{clasnature18}M. Duer \emph{et al.}, Nature, {\bf 560}, 617 (2018).
\bibitem{li18}B. A. Li, B. J. Cai, L. W. Chen and J. Xu, Prog. Part. Nucl. Phys. {\bf 99}, 29 (2018).
\bibitem{liu01}J. Y. Liu, Y. F. Yang, W. Zuo, S. J. Wang, Q. Zhao, W. J. Guo, and B. Chen, Phys. Rev. C {\bf 63}, 054612 (2001).
\bibitem{Zhang99}F. S. Zhang, L. W. Chen, Z. Y. Ming, and Z. Y. Zhu, Phys. Rev. C {\bf 60}, 064604 (1999).
\bibitem{yong20151}Hui Xue, Chang Xu, Gao-Chan Yong, Zhongzhou Ren, Phys. Lett. B {\bf 755}, 486 (2016).
\bibitem{yong20152}G. C. Yong, Phys. Rev. C {\bf 93}, 044610 (2016).
\bibitem{cpl16}F. Zhang, Chin. Phys. Lett. {\bf 33}, 012501 (2016).
\bibitem{PJA16}F. Zhang, G. C. Yong, Eur. Phys. J. A. {\bf 52}, 350 (2016).
\bibitem{lee}L. G. Sobotka, J. F. Dempsey, R. J. Charity, and P. Danielewicz, Phys. Rev. C {\bf 55}, 2109 (1997).
\bibitem{yongflow09}G. C. Yong, B. A. Li, L. W. Chen, and X. C. Zhang, Phys. Rev. C {\bf 80}, 044608 (2009).
\bibitem{jfprc86}J. Friedrich and P. G. Reinhard, Phys. Rev. C {\bf 33}, 335 (1986).
%\bibitem{ypprc13}P. Yin, J. Y. Li, P. Wang, and W. Zuo, Phys. Rev. C {\bf 87}, 014314 (2013).
\bibitem{henprc15}O. Hen, B. A. Li, W. J. Guo, L. B. Weinstein, and E. Piasetzky, Phys. Rev. C {\bf 91}, 025803 (2015).
\bibitem{yongplb17}G. C. Yong, Phys. Lett. B {\bf 765}, 104-108 (2017).
\bibitem{yongprc17}G. C. Yong and B A Li, Phys. Rev. C {\bf 96}, 064614 (2017).
\bibitem{das03}C. B. Das, S. Das Gupta, C. Gale, B. A. Li, Phys. Rev. C {\bf 67}, 034611 (2003).
\bibitem{yong18}G. C. Yong, Phys. Lett. B {\bf 776}, 447 (2018).
\bibitem{yongp4}G. C. Yong, Phys. Rev. C {\bf 96}, 044605 (2017).
\bibitem{chen98}L. W. Chen, F. S. Zhang, and G. M. Jin, Phys. Rev. C {\bf 58}, 2283 (1998).
\bibitem{shetty03}D. V. Shetty \emph{et al.}, Phys. Rev. C {\bf 68}, 054605 (2003).
\bibitem{chen032}L. W. Chen, C. M. Ko and B. A. Li, Nucl. Phys. A. {\bf 729}, 809 (2003).
\bibitem{lib99}Y. M. Zheng, C. M. Ko, B. A. Li and B. Zhang, Phys. Rev. Lett. {\bf 83}, 2534 (1999).
\bibitem{gao18}Y. Gao, G. C. Yong, L. Zhang and W. Zuo, Phys. Rev. C {\bf 97}, 014609 (2018).
\bibitem{ma01}H. Y. Zhang, Y. G. Ma, L. P. Yu, W. Q. Shen, X. Z. Cai, D. Q. Fang, P. Y. Hu, C. Zhong and D. D. Han, Chin. Phys. C {\bf 25}, 1184 (2001).
\bibitem{zhfs01}L. W. Chen, F. S. Zhang, W. F Li and Z. Y. Zhu, Chin. Phys. C {\bf 25}, 412 (2001).
\bibitem{ditorof}V. Giordano, M. Colonna, M. Di Toro, V. Greco and J. Rizzo, Phys. Rev. C {\bf 81}, 044611 (2010).
%\bibitem{liprc01}B. A. Li A.T. Sustich and B. Zhang, Phys. Rev. C {\bf 64}, 054604 (2001).
%\bibitem{jy92}J. Y. Ollitrault, Phys. Rev. D {\bf 46}, 229 (1992).
%\bibitem{jy98}J. Y. Ollitrault, Nucl. Phys. A {\bf 638}, 195c (1998).



\end{thebibliography}
\end{document}